\documentstyle[12pt]{article}

\def\AL{\frac{\alpha_f(\mu)}{ \pi}}

\def\ML{m_{\ell}^{(f)}(\mu)}
\def\MM{m_{\ell}^{(f-1)}(\mu)}
\def\O{{\cal O}}
\def \be{\begin{eqnarray}\begin{array}{rcl}\displaystyle}
\def\ee{\end{array}&&\end{eqnarray}}

\begin{document}
\begin{titlepage}
\begin{flushright}
PITHA 94/31\\
June 94
\end{flushright}
\vspace{0.8cm}
\begin{center}
{\bf\LARGE Threshold Effects on }
\vspace{0.8cm}
{\bf\LARGE the QCD Coupling \footnote{Talk given at the workshop on
"Determinations of $\alpha_s$ from inclusive observables", Aachen,
April 94.}
 $\alpha_{\overline{MS}}$}

\vspace{2cm}
{\bf Werner Bernreuther}

Institut  f\"ur Theoretische Physik\\
Physikzentrum, RWTH\\
D-52056 Aachen, Germany\\
\ \\

\vfill
{\bf Abstract:}\\
\parbox[t]{\textwidth}
{The matching condition  which determines the effect
of a heavy quark threshold on the running  of the QCD coupling
$\alpha_{\overline{MS}}$  is
reviewed. The matching scale is arbitrary to
some extent. However, this affects the value of
 $\alpha_{\overline{MS}}$
away from the threshold region only marginally.}
\end{center}
\end{titlepage}
\newpage

\section{Introduction}
The QCD coupling at the scale of the $Z$ resonance is presently obtained
 with an error of the order of about $5 \%$ (for reviews, see
\cite{Siggi}).
 At this level of precision
a careful discussion of errors, especially theoretical errors is
necessary.
When refering to $\alpha_s$
the QCD coupling defined in the $\overline{MS}$ renormalization
scheme is usually meant since most of the $\alpha_s$ determinations are
done by
comparing data with formulae obtained in this scheme. This note addresses
a topic  which is relevant in view of the precision one aims at and which
is quite often incorrectly treated in the literature:
namely, how to evolve
$\alpha_{\overline{MS}}$
which was obtained, say, at the scale of the $\tau$ mass
across heavy quark "thresholds" up to the scale of the $Z$ mass ( or vice
versa
and how to estimate the error on $\alpha_{\overline{MS}}$
associated with
threshold crossing. For the correct treatment of this problem one has to
use the matching relation for the $\overline{MS}$ coupling which was
obtained
to  order $\alpha_s^2$ \cite{Weinb} and to  order
$\alpha_s^3$ \cite{BeWe} quite some time ago.
Below the use of this
matching relation is illustrated with the example above.
\section{Matching relations for $\alpha_{\overline{MS}}$
and for the light quark masses}
As is well-known minimal subtraction renormalization
(MS) provides, apart from some
calculational advantages, a gauge- and vertex-independent, albeit
physically
unintuitive, definition of the QCD coupling. However, decoupling
of heavy
quarks in  these schemes is not manifest. In the energy region
$\mu^2 << m^2$, where $m$ is the mass of a heavy quark, the
contribution of this quark to
an observable blows up like some power of $ln(m/\mu)^2$ in a given
order of
perturbation theory.
In the case of the top quark
these logarithms can become quite large
and signal a breakdown of pertubation theory.
This behaviour is related to the fact that there is also
no decoupling of a heavy quark flavour
in the $\beta$ function which governs the scale dependence of
$\alpha_s$. In order to establish decoupling
in the MS schemes one has to resum these
logarithms. In practice this is done by matching the full $f$ flavour
theory and the effective light flavour theory in the energy region
below the heavy quark thresholds.
Consider without loss of generality QCD with $f-1$ light quarks and
one heavy
quark. In the region where the squared momenta of a certain process are
much smaller than $m^2$ the decoupling theorem
tells us that we may calculate this process also in the "light", i.e.,
the $f-1$ flavour theory up to terms of order $1/m$.
By performing such calculations both in the full and in the light theory
in the minimal subtraction scheme
one can match both theories and thereby obtain relations between
the parameters, i.e., the coupling and the light quark masses,
of both theories. (For a review, see \cite{Bern1}.) In the loop
expansion with respect to the $f$ flavour theory
these relations have the structure
\cite{BeWe,We,Bern1}:
\be
\alpha_{f-1}(\mu) = \alpha_{f-1}(\mu)
[1 + \sum_{k=1}^{\infty} C_k(x) (\AL)^k],
\label{eq:E1}
\ee

\be
\MM  = \ML [1 + \sum_{k=1}^{\infty} H_k(x) (\AL)^k] ,\cr
\ell = 1,...,f-1
\label{eq:E2}
\ee
where $\alpha_{f-1}$ and $\MM$, respectively
$\alpha_f$ and  $\ML$ denote the QCD coupling
and the running light quark masses in $f-1$ flavour, respectively
$f$ flavour
QCD for a specific MS-renormalization prescription. Furthermore $\mu$
denotes the renormalization scale,
\be
x = ln(m(\mu)/\mu)^2,
\ee
where $m(\mu)$ is the heavy quark mass defined in the $f$ flavour theory,
and $C_k(x)$ and $H_k(x)$ are polynomials in $x$ of degree $k$.
Note the following features of the matching relations (1), (2):\\
a) The polynomials $C_k$ and $H_k$ are gauge-independent and
independent of the light quark masses.\\
b) The structure of (1) and (2) is dictated by
the perturbative renormalization group; i.e.,
terms which behave like an (inverse) power in the heavy quark
mass are absent.
(In the relation between Green functions such terms are, of
course, present.) \\
c) The matching of the parameters of the
$f-1$ and $f$ flavour theories is done at a scale $\mu = \mu^*$.
This scale is arbitrary to some extent.
As usual,  $\mu^*$ should be chosen such that the perturbation
expansions can be kept under control.\\
d) For several heavy quark flavours eqs. (1) and (2) can be
applied subsequently.\\
Eq. (1) was calculated to order $\alpha_f^3$ in \cite{BeWe}. In the
following we refer to the $\overline{MS}$ scheme supplemented by
the convention that the trace of the unit matrix (in spinor space)
is kept equal to 4 in $d$ space-time dimensions.
(Note that there are additional terms in eq.(4)
if one uses another convention; see \cite{BeWe}.)  Then to this
order:
\be
\alpha_{f-1}(\mu) = \alpha_f(\mu)
[ 1 + \frac{x}{6}\AL + (\frac{x^2}{36} + \frac{11x}{24}
+\frac{7}{72}) (\AL)^2 ].
\label{eq:E4}
\ee
The criterion for the matching scale $\mu^*$, when computing
$\alpha_{f-1}$ from
$\alpha_f$ or vice versa with this formula, is that
$|x|$ must not become very much larger than one. An often used
choice is the mass of the heavy quark, $\mu^* = m(\mu=m)$. Then to
order $\alpha_f^2$:
\be
\alpha_{f-1}(m) & = & \alpha_f(m)
\label{eq:E5}
\ee
holds; whereas to order $\alpha_f^3$:
\be
\alpha_{f-1}(m) & = & \alpha_f(m) +
7 \alpha_f^3(m)/72 \pi^2 .
\label{eq:E6}
\ee
That is, the $\overline{MS}$ coupling is $not$ continuous
\footnote{Eq. (6)  yields also the relation between
the QCD scales $\Lambda_{f-1}$ and $\Lambda_f$ in the $\overline{MS}$
scheme \cite{BeWe,Marc}.} at
$\mu^* = m$. Of course one may compute from eq.(4) the matching
scale where the higher oder terms
on the right hand side of (4) cancel; but for carrying out the
matching procedure it is not necessary to obtain this value.
Instead of $\mu^* = m$ we may use
some other matching scale, for instance the threshold energy for
quark-antiquark production, $\mu^* = 2m$. Then we get from (4):
\be
\alpha_{f-1}(2m) & = & \alpha_f(2m) +
\frac{ln 4}{6} \frac{\alpha_f^2(2m)}{\pi} +
[\frac{(ln 4)^2}{36} + \frac{11 ln 4}{24} + \frac{7}{72} ]
\frac{\alpha_f^3(2m)}{\pi^2},
\label{eq:E7}
\ee
This example illustrates that one gets in general a discontinuity
already at next-to-leading order.
Note that the "continuity requirement" of the $\overline{MS}$
couplings $\alpha_{f-1}$ and $\alpha_f$ at the matching point $\mu^*$
which is often used in the literature is in general incorrect.
Nevertheless, most of the applications todate are based on
next-to-leading
order calculations. To this order eq.(5) holds which yields a
continuous coupling at $\mu^* = m$. However, for estimating the
error associated with the arbitrariness of the matching point
-- by varying $\mu^*$ in a some range around $m$  --
eq. (4) has to be used. The "continuity requirement"
overestimates this error.\\
In the $\overline{MS}$ scheme as specified above the relation
between the light quark masses in the $f-1$ and $f$ flavour
theories is given to order $\alpha_f^2$ by \cite{Bern1}:
\be
\MM &= & \ML [ 1   + \frac{1}{12}(x^2 + \frac{5x}{3}
+\frac{89}{36}) (\AL)^2 ].
\label{eq:E8}
\ee
Hence in next-to-leading order $\MM = \ML$ holds.\\
If one uses a mass-independent momentum subtraction scheme
for the definition of the QCD coupling then there is
also a non-trivial matching relation.
This relation depends on the vertex used to define
$\alpha_s$ \cite{Bern1,Bern2}. In these schemes the coupling
is discontinuous at $\mu^* = m_{\small{heavy}}$ already at
next-to-leading order.

\section{Computing $\alpha_5(m_Z)$ from $\alpha_3(m_{\tau})$ }
In order to illustrate the use of the matching relation (4)
I shall now calculate the $\overline{MS}$ coupling at the
$Z$ resonance, $\alpha_5(m_Z)$, from the recently determined
coupling $\alpha_3(m_{\tau})$ \cite{BNP,Nar94,LP,Aleph} and
determine the
error associated with the arbitrariness of the threshold
crossing points $\mu^*$. This exercise was recently also made
in \cite{Santa}. The semihadronic to electronic $\tau$ decay ratio
$R_{\tau}$ was computed in 3 flavour QCD \cite{BNP,LP}
(recall
that the perturbative contributions are known to order
$\alpha_s^3$ \cite{Russ})
and the value $\alpha_3(m_{\tau}) =0.36 \pm 0.03$ was
obtained in \cite{BNP,Nar94}.
(Alternatively
one may compute $R_{\tau}$ also in $\overline{MS}$ renormalized 4
flavour QCD with a massive $c$ quark and thereby
determine $\alpha_4(m_{\tau})$.)\\
The calculation  of  $\alpha_5(m_Z)$
from this value can be done in many different
ways. Both for $m = m_c$ and $m = m_b$ the
coefficients of the higher order terms in eq.(4) remain quite small
if we choose $\mu^*$ between these two mass values. Therefore
we may choose only one matching point, say
$\mu^* = m_{\tau}$ and compute $\alpha_4$ and $\alpha_5$ by means
of eq.(4) at this scale:
\be
\alpha_3(\mu^* = m_{\tau}) \longrightarrow
\alpha_4(\mu^* = m_{\tau}) \longrightarrow \alpha_5(\mu^* = m_{\tau})
\label{eq:E9}
\ee
and then evolve $\alpha_5(m_{\tau})$ with the 3-loop $\beta$ function
of 5 flavour QCD to $\alpha_5(\mu = m_Z)$. The solution of the
3-loop evolution equation can be represented as \cite{Marc}:
\be
\frac{1}{\alpha_f(\mu')}  =  \frac{1}{\alpha_f(\mu)}
- b_1 ln(\mu'/\mu)
-\frac{b_2}{b_1}ln(\frac{\alpha_f(\mu')}{\alpha_f(\mu)}) \cr
-\frac{1}{b_1^2}(b_3 b_1 - b_2^2)[\alpha_f(\mu') - \alpha_f(\mu)]
+ \O(\alpha_f^2)
\label{eq:E10}
\ee
with
\be
b_1  = -\frac{1}{2\pi} (11 -\frac{2f}{3})\qquad
b_2  = -\frac{1}{4\pi^2} (51 -\frac{19f}{3})\cr
b_3  = -\frac{1}{64\pi^3} (2857 -\frac{5033f}{9}
+ \frac{325f^2}{27}).
\label{eq:E11}
\ee
Using $\alpha_5(m_{\tau})$ as input in eq.(10) and iterating a few
times one gets $\alpha_5(m_Z)$.\\
Another possibility is to convert
\be
\alpha_3(\mu^* = m_{\tau}) \longrightarrow
\alpha_4(\mu^* = m_{\tau}),
\label{eq:E12}
\ee
evolve $\alpha_4$ within 4 flavour QCD to $\mu^* = m_b$, then convert
\be
\alpha_4(\mu^* = m_b) \longrightarrow
\alpha_5(\mu^* = m_b),
\label{eq:E13}
\ee
and then scale $\alpha_5$ to $\mu = m_Z$.
We shall use this procedure. As further ingredient for eq.(4) one
needs the $\overline{MS}$ masses of the $c$ and $b$ quark
at the scale $\mu^*$. We use the values given in \cite{Nar2}
which correspond to $m_c(m_{\tau}) = 1.3 \pm 0.1$ GeV and
$m_b(m_{b}) = 4.3 \pm 0.1$ GeV. The values of $m_{c,b}$ at other
scales can be obtained using the renormalization group.\\
In order to estimate the error associated with the arbitrariness of
$\mu^*$ one may first keep $\mu^* = m_{\tau}$ fixed but vary the
$b$ "threshold" between, say, 2 GeV $\leq \mu^* \leq$ 20 GeV.
Using the central value $\alpha_3(m_{\tau}) = 0.36$
as input one obtains
$\alpha_5(\mu = m_Z) = 0.123$ and this value varies by less than
$0.6\%$ when varying $\mu^*$ in the above range. Keeping the
second threshold value at $\mu^* = m_b$ and varying the $c$ threshold
between  1.7 GeV $\leq \mu^* \leq$ 4 GeV one arrives at essentially
the same result. This is in agreement with the conclusions of
\cite{Santa}. With $\alpha_3(m_{\tau}) =0.36 \pm 0.03$
one then obtains $\alpha_5(m_Z) =0.123\pm 0.004 \pm 0.001$ \newline
(cf. also \cite{Nar94}) where
the last error is a conservative estimate of the uncertainty
associated with the matching points.\\

In summary, in the evolution of
the $\overline{MS}$ coupling $\alpha_s$ across heavy quark thresholds
the matching conditon eq.(4)  comes into play.
Although the matching scales are not fixed the resulting
error on the coupling is very small as exemplified
with the above example.


\clearpage
\newpage

\end{document}